\documentclass[structabstract]{aa}  
\usepackage{graphicx}
\usepackage{natbib}
\usepackage{longtable}
\begin{document}

   \title{Starspots on the fastest rotators in the $\beta$ Pic moving group}

   \author{D. Garc\'{\i}a-Alvarez\inst{1,2,3}
          \and
	  A.F. Lanza\inst{4}
	  \and
	  S. Messina\inst{4}
	  \and
          J.J. Drake\inst{3}
	  \and
	  F. van Wyk\inst{5}
	  \and
	  R.R. Shobbrook\inst{6}
	  \and
	  C.J. Butler\inst{7}
	  \and
	  D. Kilkenny\inst{8}
	  \and
	  J.G. Doyle\inst{7}
	  \and
	  V.L. Kashyap\inst{3}}

   \institute{Instituto de Astrof\'{\i}sica de Canarias, E-38205 La Laguna, Tenerife, Spain
	 \and	 
	     Grantecan CALP, 38712 Bre{\~na} Baja, La Palma, Spain
	 \and
	     Harvard-Smithsonian CfA, 60 Garden Street, Cambridge, MA 02138, USA
	 \and
	     INAF - Osservatorio Astrofisico di Catania, via S. Sofia 78, 95123 Catania, Italy
         \and
	     South African Astronomical Observatory, P.O. Box 9, Observatory 7935, Cape Town, South Africa
	 \and
	     Research School of Astronomy and Astrophysics, Australian National University, Canberra, ACT, Australia
	 \and
	     Armagh Observatory, College Hill, Armagh BT61 9DG Northern Ireland
	 \and
	     Department of Physics, University of the Western Cape, Private Bag X17, Bellville 7535, South Africa. 
	}

   \date{Received ; accepted }

 
  \abstract
   {}
  { We carried out high-resolution spectroscopy and {\it BV(I)}$_C$ photometric
monitoring of the two fastest late-type rotators in the nearby $\beta$
Pictoris moving group, HD~199143 (F7V) and CD$-$64$^{\circ}$1208 (K7V). The motivation for this 
work is to investigate the rotation periods and photospheric spot 
patterns of these very young stars, with a longer term view to 
probing the evolution of rotation and magnetic activity during the early
phases of main-sequence evolution. We also aim to derive information on key physical parameters, such as rotational velocity and rotation period.}
   { We applied maximum entropy (ME) and
Tikhonov regularizing (TR) criteria to derive the surface spot map distributions of the optical modulation observed in HD~199143 (F7~V) and CD$-$64$^{\circ}$1208 (K7~V). We also used cross-correlation techniques to determine stellar parameters such as radial velocities and rotational velocities. Lomb-Scargle periodograms were used to obtain the rotational periods from differential magnitude time series.}
   { We find periods and inclinations of 0.356 days and
21.5~deg for HD~199143, and 0.355 days and 50.1~deg for CD$-$64$^{\circ}$1208. The spot maps of HD~199143 obtained
from the ME and TR methods are very similar, although the latter gives a
smoother distribution of the filling factor.  Maps obtained at two
different epochs three weeks apart show a remarkable increase in spot
coverage amounting to $\sim 7$\% of the surface of the photosphere
over a time period of only $\sim 20$ days. The spot maps of CD$-$64$^{\circ}$1208
from the two methods show good longitudinal agreement, whereas the
latitude range of the spots is extended to cover the whole visible
hemisphere in the TR map. The distributions obtained from the first light curve of HD~199143 show the presence of an extended and asymmetric active longitude
with the maximum filling factor at longitude $\sim 325^{\circ}$. A
secondary active longitude is present at $\sim 100^{\circ}$. The spotted area distributions on CD$-$64$^{\circ}$1208 show two
active longitudes separated by about $180^{\circ}$, which is not
unusual on such very active stars. 

}
  {}

   \keywords{ Stars: rotation, Stars: late-type, Stars: individual: HD~199143, CD$-$64$^{\circ}$1208, Stars: activity, Stars: starspots}
   \maketitle

%

\section{Introduction}

Moving groups (MGs) are kinematically coherent groups of stars that
probably share a common origin---the evaporation of an open cluster,
the remnants of a star formation region, or a juxtaposition of several
small star formation bursts at different epochs in adjacent cells of
the velocity field. Recent years have seen the discovery of several
MGs within 100\,pc of the Sun (e.g.; TW Hydrae, Horologium-Tucana
and $\beta$ Pictoris). Stars in these groups range between millions
and tens of million years in age, and their study should tell us
about the formation and evolution of sparse stellar associations. The
stars themselves provide a rare glimpse of the very earliest
main-sequence evolution and an opportunity to study magnetic activity
at its extremes. Young late-type fast rotating members of MGs offer a 
unique laboratory for studying stellar dynamos, magnetic structures and 
coronal heating.

The $\beta$ Pictoris moving group (BPMG) is one of the youngest
\citep[$\sim$12\,Myr,][]{kai04} and the closest MGs to the Sun 
\citep[$\sim$36\,pc,][]{zuc01}. \cite{zuc01} and \cite{son03}
 report a total of 35 known members of the BPMG,
each with one or more characteristics indicative of extreme
youth. \cite{kai04} found a high lithium abundance in one of
the BPMG members.  The best-known members of the BPMG are
early-type stars, whereas few studies have concentrated on the
late-type stars representing three fourths of the known BPMG membership.
The rotational properties of the late-type members have been
recently investigated by \cite{messina10} as part 
of the RACE-OC project.

We have carried out high resolution spectroscopy and {\it BV(I)$_C$}
photometric monitoring of a selected sample of young late-type stars
(spectral type later than F2). The sample chosen for this work has
been selected from previously established members of the BPMG, based on
photometric and kinematic properties. The motivation for this 
work is to investigate the rotation periods and photospheric spot 
patterns of these very young stars, with a longer-term view to
probing the evolution of rotation and magnetic activity during the early
phases of main-sequence evolution. 

Here, we present results for two members of our sample, HD~199143 and
CD$-$64$^{\circ}$1208. These are the fastest rotators among the known late-type
stars in the BPMG. HD~199143 has a $v\sin i$ of 128 km\,$s^{-1}$
\citep{tor06} which is the largest of any known single 
solar-like star within 50~pc. 
\cite{anc00} proposed that 
HD 199143 has been spun up by accretion of material from a
close T Tauri-like companion responsible for the emission lines,
the ultraviolet variability and the excess infrared emission. 
In the case of CD$-$64$^{\circ}$1208, $v \sin i > 100 $~km~s$^{-1}$
\citep{zuc01} and its unusually strong Li for a 10 Myr old star has been
postulated as possibly being due to its very fast rotation \citep{sod93}. 
Both stars represent a special opportunity to study the stellar atmospheric
activity in young solar-like ultra-fast rotators. 

{ Further information on the targets and observations are given in
Sect.~\ref{sec:targets} and Sect.~\ref{sec:obs} respectively. Stellar parameters, time series analysis and phasing of the light curves are
described in Sect.~\ref{sec:parameters}, Sect.~\ref{sec:timeseries} and Sect.~\ref{sec:phasinglightcurves} respectively. The spot map distributions are shown in Sect.4~ref{sec:spotmap}, while in Sect.~\ref{sec:results} we present the results. 
The conclusions are given in Sect.~\ref{sec:conclusions}.}

\section{HD~199143 and CD$-$64$^{\circ}$1208}
\label{sec:targets}

As members of the young $\beta$ Pictoris moving group, the F7 dwarf
HD~199143 ($d=47.7$~pc) and the K8 dwarf CD$-$64$^{\circ}$1208 ($d=29.2$~pc) probe
the immediate post-T~Tauri phase, during which stars contract to the
zero-age main sequence.  HD~199143, is a wide (1\arcmin) binary system.  It
has the largest $v \sin i$ of any solar-like star known within 50 pc
\cite[$v \sin i=128$~km~s$^{-1}$,][]{tor06}.  HD 199143 forms a
physical pair with the Li-rich late-type dwarf BD$-$176128.  
It exhibits strong chromospheric
activity in H$\alpha$ and in the ultraviolet \citep{anc00}. 
The metallicity of HD~199143, as estimated from photometric indices
in the survey of \cite{nor04}, is [Fe/H]$ = -0.07$.  Its
logarithmic X-ray luminosity (in erg~s$^{-1}$) as measured with ROSAT
is $\log L_{\rm x} = 30.56$, but $\log (L_{{\rm x}}/L_{{\rm bol}})$ is
-3.3 \citep{zuc01}, which puts the star below  saturation \citep{jef05}. 
If the star has a short period ($<$0.5 days), it would be placed in the supersaturation range \citep{gar08}. 

{ CD$-$64$^{\circ}$1208 is a triple system consisting of a visual binary (whose primary has a  K8~V  spectral type 
as estimated from {\it BV(RI)$_{C}$} photometry \citep{zuc01} with a separation of 0.18~arcsec
 and $\Delta K =2.3$~mag \citep{chauvin10} and a wide A7~V companion at a distance of 70~arcsec (HIP~92024).} A spectrum obtained with the Double Beam Spectrograph (DBS)
at the 2.3 m telescope in SSO \citep{zuc01} {is dominated by the K8~V component and } shows strong
H$\alpha$ in emission (EW $\sim$ 2.2 \AA), a strong Li 6708\AA~feature
(EW $\sim$ 580~m\AA) and fast rotation.  The unusually strong Li signature 
is possibly due to its very fast rotation \citep[e.g. see][]{sod93}.

{ In principle, the dilution of the optical flux of CD$-$64$^{\circ}$1208 by its close visual companion may affect its light variations and have an impact on the spot modelling. If the companion is a late-type active star, say later than F5~V, and rotates with a period commensurable with that of the target star, its rotational modulation may produce a spurious additional signal in phase with the light variation of  CD$-$64$^{\circ}$1208 that will introduce a systematic error on the derived distribution of its spotted area vs.\ longitude. An upper limit for this error can be estimated by assuming that the $V-K$ colour of the visual companion is at least $1.10$~mag, corresponding to a star of spectral type F5~V \citep[cf. ][ Sect.~7.5]{Cox00}, and that its optical light modulation has an amplitude of $0.2$~mag, i.e.\ comparable with that of CD$-$64$^{\circ}$1208. Thus, the upper limit to the relative optical flux variation of the companion is $\Delta F_{\rm V}/F_{\rm V} = 8.8 \times 10^{-3}$, which implies a maximum systematic error in the evaluation of the total spotted area of the target of $\sim 0.9$ per cent of the star disc, on the hypothesis of completely dark spots. Since the companion is likely to be significantly cooler than F5~V, the actual error may be considerably smaller and may be safely neglected, given that the amplitude of the light variation of the target is $\sim 0.2$~mag in the {\it V} passband, corresponding to a maximum spotted area of  $\sim 17$ per cent of the star's disc (cf. Sect.~\ref{sec:obs}).}

\section{Observations} 
\label{sec:obs}
\subsection{Photometry}

\begin{figure}[here]
   \centering
   \includegraphics[angle=90,width=9cm]{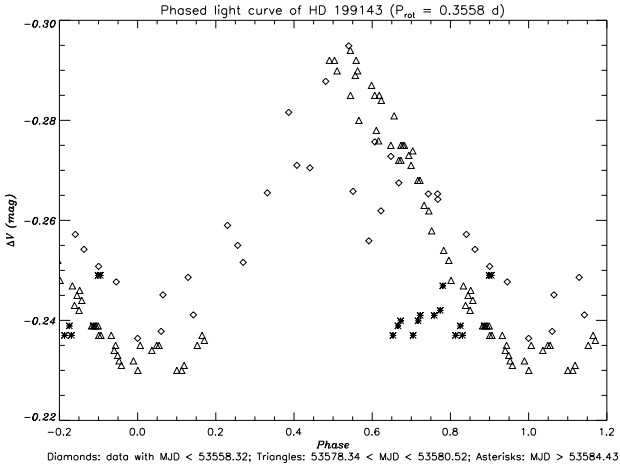}
      \caption{The V-band differential magnitude of HD~199143 versus the phase 
as given by Eq.~\ref{ephemF8V}. Different symbols refer to different epochs
of observation as indicated (MJD=HJD$-$24500000). }
         \label{lightcurve_hd199143}
   \end{figure}

\begin{figure}[here]
   \centering
   \includegraphics[angle=90,width=9cm]{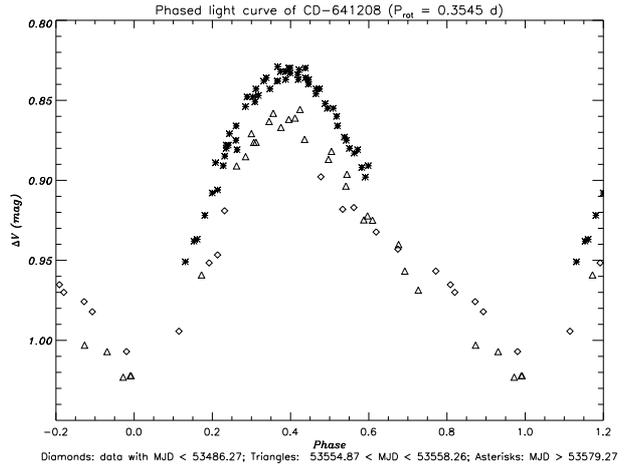}
      \caption{The V-band differential magnitude of CD$-$64$^{\circ}$1208 
versus the phase as given by  Eq.~\ref{ephemK8V}. Different symbols refer to different observational
epochs as indicated (MJD=HJD$-$24500000).}
         \label{lightcurve_cd641208}
   \end{figure}

Photometric observations of the stars HD~199143 and CD$-$64$^{\circ}$1208 were
obtained with Cousins {\it BV(I)$_C$} filters 
during 2005 using telescopes at the South African
Astronomical Observatory (SAAO) and the Siding Spring Observatory
(SSO), Australia, and are summarised in Table~\ref{log_photometry}.  
SAAO observations were made using the Modular Photometer on the
0.5 m telescope at Sutherland.  The observations were made with
reference to the Cousins {\it UBV(RI)$_C$} standards in the E-regions \citep{men89}. 
SSO observations employed the 0.6-m reflector and PMT
detector.  The measurements of targets, HD~199143 and
CD$-$64$^{\circ}$1208, were alternated with two comparison stars. The light curves
of HD~199143 and CD$-$64$^{\circ}$1208 are shown in Fig.~\ref{lightcurve_hd199143}
and Fig.~\ref{lightcurve_cd641208} respectively.

\begin{table}
\begin{center}
\caption{\small{The log of the {\it BV(I)${_C}$} photometric observations of the targets from SAAO and SSO.}}
{
\begin{tabular}{@{}lccc@{}}
\hline
\hline
Object & HJD & Site & No. of  \\
       & 2453000.0+ &  & datapoints\\
\hline
CD$-$64$^{\circ}$1208 & 473.14-486.27 & SSO & 16 \\
CD$-$64$^{\circ}$1208 & 554.88-558.26 & SSO & 27\\
CD$-$64$^{\circ}$1208 & 579.28-584.38 & SAAO & 57 \\
HD~199143 & 555.02-558.32 & SSO & 27 \\
HD~199143 & 578.34-584.52 & SAAO & 78 \\
\hline
\end{tabular}
}
\label{log_photometry}
\end{center}
\end{table}
\begin{table*}
\begin{center}
\caption{\small{The log of spectroscopic observations of the targets and standard stars from SAAO.}}
{
\begin{tabular}{@{}lccrcl@{}}
\hline
\hline
Object & \multicolumn{2}{c}{UT Start} & Exp.time & No. of & Comments \\
       &&  & (s) & frames&\\
\hline
CD$-$64$^{\circ}$1208 &2004 Aug 03 & 21:37 & 1800 & 1 & Target Star\\
\hline
CD$-$64$^{\circ}$1208 &2004 Aug 04 & 20:51 & 1800 & 1 & Target Star\\
HD~199143 & 2004 Aug 04 &23:41 & 1800 & 1 & Target Star\\
HD 4247   & 2004 Aug 04 &02:12 & 600 & 4 & F0\,V template\\
HD 886    & 2004 Aug 04 &02:40 & 250 & 1 & B2\,IV telluric std\\
\hline
HD 142764 & 2004 Aug 06 &19:40 & 300 & 2 & K5\,V template\\
HD 171391 & 2004 Aug 06 &20:32 & 1200 & 5 & G8\,III radial velocity std\\
\hline
HD 160032 & 2004 Aug 07 &17:07 & 600 & 2 & F4\,V template\\
HD~199143 & 2004 Aug 07 &19:36 & 1200 & 1 & Target Star\\
HD~199143 & 2004 Aug 07 &20:07 & 1200 & 1 & Target Star\\
\hline
HD 160032 & 2004 Aug 08 &18:10 & 500 & 2 & F4\,V template\\
\hline
\end{tabular}
}
\label{log_spectroscopy}
\end{center}
\end{table*}
\begin{table*}
\begin{center}
\caption[]{Stellar parameters}
\small
\begin{tabular}{ l c c c c r r r r }
\noalign{\smallskip}
\hline
\hline
\noalign{\smallskip}
Object &   {$B$--$V$}& d&
$v\sin{i}$ & $V_{\rm hel} \pm \sigma_{\rm V_{hel}}$ &
$U\pm \sigma_{\rm U}$ & $V \pm \sigma_{\rm V}$ & $W \pm \sigma_{\rm W}$ \\
              &          &(pc)&(km s$^{-1}$)
& (km~s$^{-1}$) & (km~s$^{-1}$) & (km~s$^{-1}$) & (km~s$^{-1}$) \\
\noalign{\smallskip}
\hline
\noalign{\smallskip}
HD~199143    &  0.48& 47.7&115.5$\pm$7.5&   2.5$\pm$2.9 & $-$10.57$\pm$1.77 & $-$15.83$\pm$1.03&   $-$9.45$\pm$0.90\\
CD-64$^{\circ}$1208    & 1.2& 29.2&121.3$\pm$15.3&   2.2$\pm$2.2 & $-$3.06$\pm$2.06 & $-$11.10$\pm$1.33&   $-$15.63$\pm$1.80\\
\noalign{\smallskip}
\hline
\noalign{\smallskip}
%
%
\end{tabular}
\label{tab:par}
\end{center}
\end{table*}

\subsection{Spectroscopy}

High resolution spectra of HD~199143 and CD$-$64$^{\circ}$1208 were obtained
during August 2004 at SAAO, using the GIRAFFE fibre-fed echelle
spectrograph attached to the 1.9-m Radcliffe telescope.  This
instrument is a copy of the MUSICOS spectrograph currently in use on
the Bernard Lyot telescope at Pic du Midi Observatory, France
\citep{bau92}. The resolving power is $\lambda/\Delta\lambda 
\approx 42 000$,
corresponding to a width of 2 pixels of the $1024 \times 1024$ TEK
CCD. A Th-Ar arc lamp was used for wavelength calibration. Arc spectra
were taken at regular intervals to calibrate possible  
drifts. Flat-fielding was performed by taking the spectrum of a
tungsten lamp and also by illuminating the CCD with uniform light
through a diffusing screen. The wavelength range was 4300--6680~\AA~
spread over 52 orders. Table~\ref{log_spectroscopy} gives the log of
the observations.

\begin{figure*}[ht!]
   \centering
   \includegraphics[angle=90,width=\textwidth]{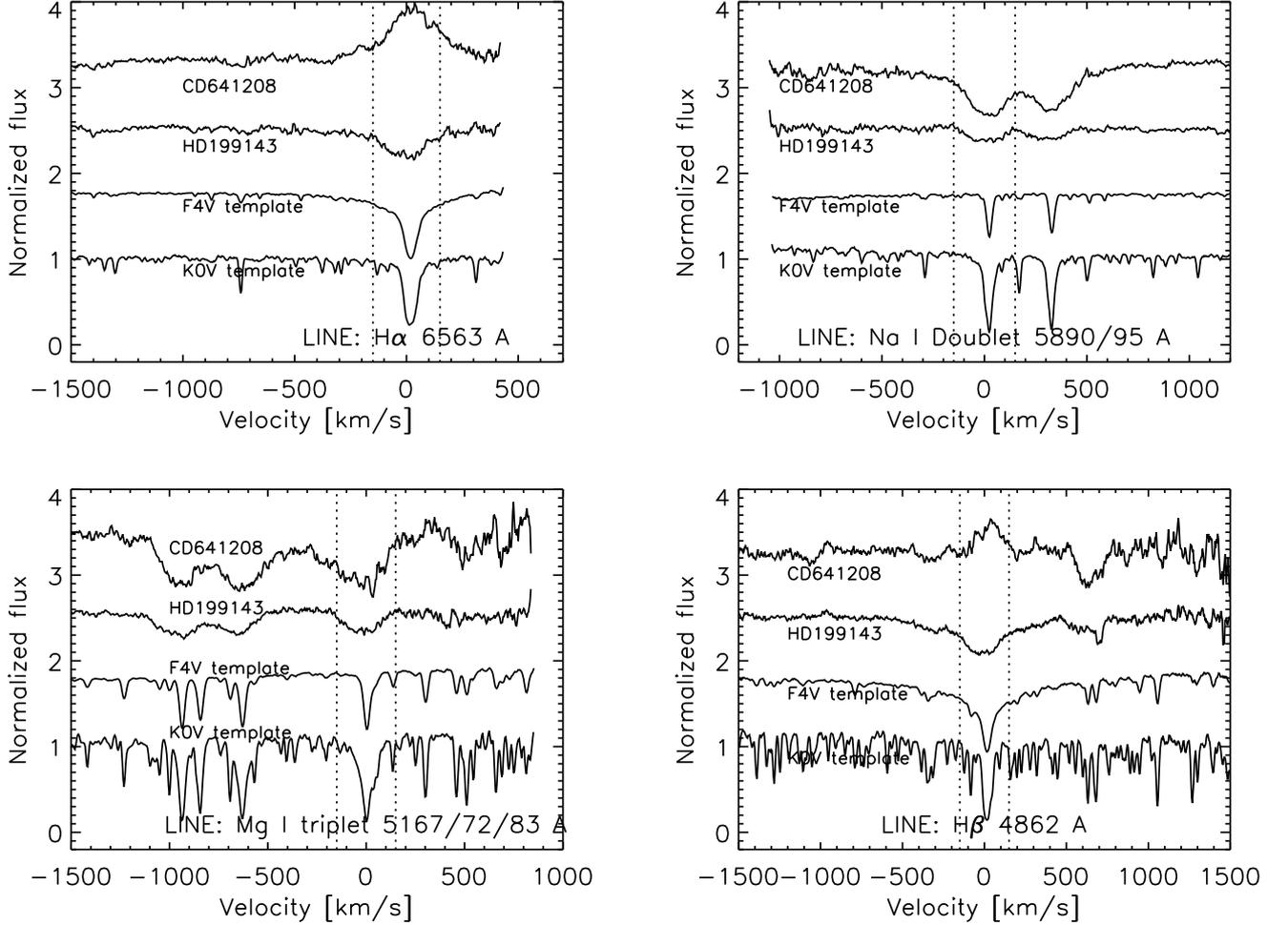}
      \caption{High-resolution spectra of HD~199143 and CD$-$64$^{\circ}$1208. For comparison, we also show the spectra of template stars with similar spectral types.}
         \label{spectra_hd199143}
   \end{figure*}

The spectra were reduced in a standard fashion, after which the
continuum was normalized to unity. Spectral regions centred on 
some prominent lines of interest are shown in
Fig.~\ref{spectra_hd199143}, together with the spectra of stellar 
``templates'' of similar
spectral type obtained during the same campaign. 
Highly broadened (FWHM $>$ 250~km~s$^{-1}$) photospheric absorption lines
(most prominently H$\alpha$, the Mg~I triplet, the Na~I doublet and
H$\beta$) are visible in the spectra of both HD~199143 and
CD$-$64$^{\circ}$1208. The same lines are present in the template stellar spectra,
but are, as expected, much narrower.

   \begin{figure}[here]
   \centering
   \includegraphics[angle=90,width=9cm]{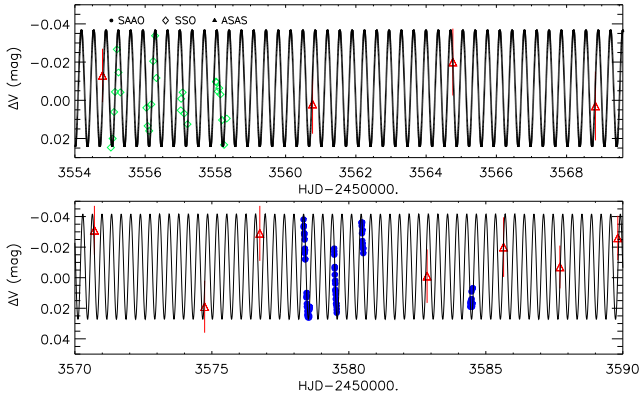}
      \caption{The V-band differential magnitude of HD~199143 versus time
with overplotted a sinusoidal fit with the rotation period of $0.3558$~d. Different symbols refer to different observatories, as labelled.}
         \label{HD199143_series}
   \end{figure}


\section{Stellar parameters}
\label{sec:parameters}
\subsection{Radial velocity and space motions}

Heliocentric radial velocities were determined using the 
cross-correlation technique.
The spectra of each star were cross-correlated order by order,
using the routine {\sc fxcor} in IRAF, against spectra of radial velocity
standards of similar spectral types 
(see Table~\ref{log_spectroscopy}).
For each order, the velocity is derived 
from the position of the peak of the cross-correlation function (CCF) 
found by fitting a Gaussian to the top of the function.
Radial velocity errors are computed by {\sc fxcor} based on the 
fitted peak height and the antisymmetric noise as described by 
\cite{ton79}.
The radial velocities calculated for each order are weighted by their 
errors, and a mean value is obtained for each observation.
Orders including chromospheric features and prominent telluric lines
were excluded when determining this mean velocity.
Finally, a weighted mean radial velocity is determined using all the 
observations from all the observing runs. 
In Table~\ref{tab:par} we list the average
heliocentric radial velocity ({\it V}$_{\rm hel}$)
and its associated error ($\sigma_{\rm V_{hel}}$). The final mean radial velocities we have obtained for HD~199143 and CD$-$64$^{\circ}$1208
are $v_{\rm r} = 2.5$ and $v_{\rm r} = 2.2$ km s$^{-1}$, respectively.\\

We have used these mean radial velocities, together with the spectroscopic parallaxes 
({\it d} = 47.7 and 29.2 pc) and proper motions from the Tycho-2 Catalogue
\citep{hog00}, 
to calculate the Galactic space-velocity components ($U$, $V$, $W$) 
in a right-handed coordinated system (positive in the directions of the
Galactic center, Galactic rotation, and the
North Galactic Pole, respectively), as determined 
by \cite{mon01}. The resultant values are given in Table~\ref{tab:par}.
The positions of HD~199143 and CD$-$64$^{\circ}$1208 in the Galaxy velocity diagram 
confirm their membership of the BPMG. These values are similar to those reported by \citep{zuc01}.
\subsection{Rotational velocity}
To determine an accurate rotational velocity for HD~199143 and
CD$-$64$^{\circ}$1208, we again make use of the IRAF {\sc fxcor} cross-correlation
routine.  For each of the observing runs, the observed spectra of
HD~199143 and CD$-$64$^{\circ}$1208 were cross-correlated against the spectrum of
the template star and the width (FWHM) of the cross-correlation function (CCF) determined. The
calibration of this width to yield an estimate of $v\sin{i}$ is done
by cross-correlating artificially broadened spectra of the template
star with the original template star spectrum.  Broadened template
spectra are created for $v\sin{i}$ spanning the expected range of
values by convolution with a theoretical rotational profile using 
the program {\sc starmod}.  The resultant relationship
between $v\sin{i}$ and FWHM of the CCF was then fitted with a
fourth-order polynomial.  We have tested this method with stars of
known rotational velocity and similar spectral type to HD~199143 and CD$-$64$^{\circ}$1208. 
On average, we have obtained values with errors of about 5\% compared 
with those from the literature. The
uncertainties in the $v\sin{i}$ values obtained by this method have
been calculated using the parameter $R$ defined by Tonry \& Davis
(1979) as the ratio of the CCF height to the RMS antisymmetric
component. This parameter is computed by the IRAF task {\sc fxcor} and
provides a measure of the signal-to-noise ratio of the CCF. \cite{ton79} 
show that errors in the FWHM of the CCF are proportional
to $(1 + R)^{-1}$, and \cite{har86} and \cite{rho01} 
found that the quantity $\pm v\sin{i}(1 + R)^{-1}$ provides a good 
estimate for the 90$\%$ confidence level of a $v\sin{i}$ measurement.
Thus, we have adopted $\pm v\sin{i}(1 + R)^{-1}$ as a reasonable
estimate of the uncertainties in our $v\sin{i}$ measurements.

We have determined $v\sin{i}$ by this method for all the spectra
available for HD~199143 and CD$-$64$^{\circ}$1208.
We found $v\sin{i}$ values that match within the errors by using spectra from different nights
and at different epochs. We note that $v\sin{i}$ derived by this method
is not sensitive to possible asymmetries of the CCF caused by 
starspots: such changes are mainly in the peak of the CCF, whereas 
the $v\sin{i}$ determination is based on the FWHM.
The resulting error-weighted means 
for all the observing runs for HD~199143 and CD$-$64$^{\circ}$1208 are $115.5 \pm 7.5$ and $121.3 \pm 15.3$ km~s$^{\rm -1}$, which are the values given in Table~\ref{tab:par}. These values 
are very similar to those given by \cite{tor06}.  
\cite{wei10} obtained a larger value of 
$v\sin~i$ = 161$\pm$12 km s$^{\rm -1}$ for HD~199143. They cross-correlated 
the stellar spectra (3900-6800 \AA) with appropriate numerical templates for 
the respective stellar spectral type \citep[see][]{bar79}.

\section{Time series analysis: the search for rotation periods}
\label{sec:timeseries}

{The spectral types and measured $v \sin i$ values confirm that 
HD~199143 and CD$-$64$^{\circ}$1208 are fast rotators with periods $P_{\rm rot} < 1$
days.  We therefore limited the search for periodicity in 
the V-band differential photometry to the range $0.1-10.0$~days to allow  for possible modulations on shorter as well as  longer periods.  
We applied the Lomb-Scargle periodogram \citep{scargle82} for unevenly spaced data 
following the prescriptions of \cite{hor86}.

For HD~199143,  the Lomb-Scargle periodogram shows the
maximum power at a period of 0.3558 days with a false-alarm
probability smaller than $10^{-6}$. The formal error on such a period
is 0.004 days as derived from the FWHM of the corresponding
periodogram peak, and is determined by the limited
time extension of the data set (only 29.5~days). However,  putting
the V-band magnitude in phase with the 0.3558-d period, we find that a
period change as small as 0.001~days gives rise to a significant
increase in the dispersion of the light curve points. Therefore, we
prefer to assume the latter as the most probable error of the period
determined.

Visual inspection of the differential magnitude time series (see Fig.\,\ref{HD199143_series}) 
shows how good  the agreement is between
the data and the sinusoidal fit with the 0.3558-d rotation period. We have retrieved  contemporary V-band observations of HD\,199143 from the ASAS  archive (All Sky Automated Survey, \citep{pojmansli97} and these are plotted together 
with our own observations. Also, the ASAS data agree 
with the period we found. We have searched for the rotation period also in the 
ASAS time series. However, owing to the low precision of HD\,199143, ASAS 
photometry ($\sim$0.035 mag) and the low amplitude of the rotational modulation 
($\sim$ 0.055 mag), our period search  did not
allow us to derive any significant periodicity.

For CD$-$64$^{\circ}$1208, the Lomb-Scargle periodogram exibits two peaks of comparable power and 
significance level greater than 99\%, one at {\it P}~=~0.3545d and another at {\it P}~=~0.5459d.
However, only the shorter period, when combined with the estimated stellar radius ({\it R}~=~1.1 {\it R}$_{\odot}$),
 gives an equatorial velocity consistent with the measured $v\sin i$~=~121.3 km~s$^{-1}$.
Moreover, our rotation period estimate is confirmed by the {\it P}~=~0.345 $\pm$ 0.04~d value as determined by
\citep{messina10} making use of the ASAS photometry. In fact, in the case of CD$-$64$^{\circ}$1208,
although the ASAS photometry has a  precision similar to that of HD~199143,  the rotational modulation is 
a factor 3 larger ($\sim$ 0.16 mag), thus allowing them safely to  establish the rotation period. A significant periodicity was  found in 7 out of the 12 time segments in which the entire ASAS  time series was sectioned.
The error in our derived period can be assumed to be of the order of 0.001~days because a
larger uncertainty significantly increases  the dispersion of the light
curve points. On the other hand, the periodicity at $0.5459$~d may be the result of  aliasing and noise fluctuations affecting our time series.}

\section{Phasing of the light curves}
\label{sec:phasinglightcurves}

The observations of HD~199143 are phased with the
ephemeris:
\begin{equation}
{\rm HJD}_{\rm min} = 2453555.02423 + 0.3558 \times E,   
\label{ephemF8V}
\end{equation}
giving the epoch of the minimum light. The V-band light curve is shown
in Fig.~\ref{lightcurve_hd199143}, where different symbols refer to
observations collected at different epochs in order to show how the
shape of the light modulation changes vs.\ time. Specifically, three
intervals can be identified during which the light curve remains
stable, i.e.\ from HJD~2453555.02 to 2453558.32 (diamonds), from
2453578.34 to 2453580.52 (triangles) and from 2453584.43 to 2453584.52
(asterisks). The photometric points belonging to the first time 
interval sample the entire light curve, whereas those of the second
interval show a gap between phases 0.2 and 0.5.  A variation of the
peak amplitude of the light curve is clearly evident between the two
time intervals, making it impossible to put the data together to
improve the phase sampling of the light curve. The third data set does
not show any significant variation and covers only the phases between
0.75 and 0.80: these points will not be considered for the subsequent
spot modelling.  The brightest differential {\it V} magnitude along the
whole time series is -0.295 and was assumed as the unspotted
magnitude for the spot modelling of HD~199143. An inspection of the
ASAS magnitudes - which better sample the range of variability because of
their 9 year baseline - confirms that our choice of unspotted value is the
correct one.

The observations of CD$-$64$^{\circ}$1208 are phased with the ephemeris:
\begin{equation}
{\rm HJD}_{\rm min} = 2453473.20700 + 0.3545 \times E,   
\label{ephemK8V}
\end{equation}
The V-band light curve is plotted in Fig.~\ref{lightcurve_cd641208},
with different symbols indicating data collected at different times.
Also in this case, three time intervals can be identified during which
the light modulation remained more or less stable, i.e.\ from
2453473.14 to 2453486.27 (diamonds), from 2453554.87 to 2453558.26
(triangles) and from 2453579.27 to 2453584.37 (asterisks).  The
points of the first two data sets trace very similar light
modulations, except for the phase interval between 0.7 and
1.0. However, they were considered together in order to improve the
phase sampling of the combined light curve that was fitted with
our spot model. The third data set consists of points between phases
0.15 and 0.6 that are systematically brighter than the corresponding
points of the first and the second data sets. Since the
systematic difference is about 0.025 mag, the most likely explanation
is a decrease in the area of the component of the spot pattern  uniformly
distributed over the star that made it brighter. A change of
the area of the order of $2-3$\% of the star disc over a time interval
of about 20 days is enough to account for such a variation and cannot
be considered unusual for such a very active star.

The brightest differential {\it V} magnitude throughout the whole time series is
0.829 and this was assumed as the unspotted magnitude for the spot modelling of
CD$-$64$^{\circ}$1208. As done in the previous case, we found that the correctness of this value is 
confirmed  by the ASAS magnitudes timeseries.

\section{Spot map distributions}
\label{sec:spotmap}
\subsection{Spot modelling technique}
\label{sec:spot}

The classic Doppler Imaging technique can provide us with a map of the
stellar photosphere of an active star \citep[e.g.,][]{str02,strassmeier09} but its
applicability is strongly limited in the present case owing to the
extreme rotational broadening of the photospheric spectral
lines. Therefore, spot modelling based on wide-band light curve
fitting appears to be more appropriate for a preliminary
characterization of the photospheric activity of our stars.

The reconstruction of the brightness distribution over the surface of
a star by modelling the rotational modulation of wide-band optical
fluxes is an ill-posed problem. The modulation provides information
only on the variation of the projected spot area {\it versus} the rotation 
phase, i.e., the stellar longitude.  In principle, it is possible to
determine a unique spot distribution by minimizing the $\chi^{2}$ of
the residuals between the observed and the model light curve, but such
an approach is unsatisfactory because the solution is unstable, i.e.\
small variations in the input data lead to large changes in the spot
map. This is due to the role played by the noise in the $\chi^{2}$
minimization, implying that most of the structure appearing on the
spot map actually comes from the overfitting of the light fluctuations
produced by  intrinsic stellar variability or measurement errors.

It is possible to overcome the uniqueness and stability problems by
introducing a regularizing function into the solution process. This
corresponds to the {\it a priori} assumption of some specific statistical
properties for the spot map, which allows us to select one stable map
among the potentially infinite  maps that can fit a given
light curve.  The two most used a priori assumptions are the maximum
entropy \citep[hereinafter ME;][]{gul84,vog87} and
the Tikhonov \citep[hereinafter TR;][]{pis90}
regularizations. These are particularly well-suited when the spot map
consists of an array that specifies the spot covering factor in each
surface element  of the model star.  The covering factor $f$
gives the specific intensity $I$ of each surface element according to
the definition: $I= (1-f) I_{\rm u} + f I_{\rm s}$, where $I_{\rm u}$
is the specific intensity of the unspotted photosphere of the surface
element, $I_{\rm s}$ that of the spotted photosphere, and $0<f<1$.

A detailed description of the application of the ME and TR
regularizations to spot modelling problems was presented by \cite{lan98} 
to whom we refer the reader for a detailed discussion of
our approach.  Here we only recall that the ME spot maps are computed
by a constrained minimization of a functional $Q_{\rm ME}$, 
\begin{equation}
Q_{\rm ME}= \chi^{2} -\lambda_{\rm ME} S, 
\end{equation}
whereas the TR maps are computed by the constrained minimization of 
\begin{equation}
Q_{\rm TR} = \chi^{2} + \lambda_{\rm TR} TR.
\end{equation}
Here, $\chi^{2}$ is the normalized sum of the squared residuals
between the observed and the synthesized light curves, $S$ is the entropy of the spot map, $TR$
is the Tikhonov functional of the spot map, and $\lambda_{\rm ME} > 0$ and
$\lambda_{\rm TR} > 0$ are the Lagrange multipliers for the ME and the TR
regularizations, respectively.  The explicit expressions for $S$ and
$TR$, and the procedure of evaluation of the Lagrange multipliers are
described by \cite{lan98}.

The synthesized light curve for a given distribution of the covering
factor is computed by adopting a spheroidal geometry for the star, i.e.\
it is represented as an ellipsoid of semi-axes $a=b$ and $c<a$, whose
flattening is due to the centrifugal potential produced by the rapid
rotation. We compute the flattening in the Roche approximation as
\begin{equation}
\frac{c}{a} = 1 - \frac{1}{2} \left( \frac{\Omega^{2} a^{3}}{G M}\right),
\end{equation}
where $c$ is the polar radius, $a$ the equatorial radius, $\Omega$ the
angular velocity of rotation, $M$ the mass of the star and $G$ the
gravitation constant. The effective temperature $T$ on the photosphere
of a distorted star depends on the local gravity \citep[e.g.,][]{kal99}. 
For a star having an outer convective envelope, a good
approximation is given by
\begin{equation}
T = T_{\rm p} \left[ 1 + 0.08 \times \left( \frac{g-g_{\rm p}}{g_{\rm p}} 
\right) \right],
\label{gravdark}
\end{equation}
where $T_{\rm p}$ is the effective temperature at the pole, $g$ is the
local surface gravity and $g_{\rm p}$ is the gravity at the pole. By
considering such a gravity darkening effect, we compute the
distribution of the effective temperature on the surface of the star
and adopt \cite{kur00} models to compute the brightness emerging
from each surface element. Adding the effect of the
limb-darkening, treated in the linear approximation, we derive the
unperturbed brightness $I_{u}$ over the stellar photosphere. The
brightness of the spotted photosphere is given by: $I_{\rm s} = C_{\rm
s} I_{\rm u}$, where the contrast $C_{\rm s}$ depends on the spot
effective temperature and is assumed  constant over the
photosphere of the star. In order to compute the emerging flux with a
relative precision of the order of $10^{-4}$, the stellar photosphere
is subdivided into squared elements of side $s_{\rm e} = 1^{\circ}$,
the fluxes of which are summed to compute the total flux from the
stellar disc.

Absolute properties of the spots cannot
be extracted from single-band data because systematic and other errors arise
 from the unknown unspotted light level of the star and
the assumption of single-temperature spots.  Specifically, the
brightest magnitude observed in such active stars is probably fainter 
than the true unspotted magnitude because spots may always be present
on their photospheres. However, our model can be applied to derive a
map of the component of the spot pattern unevenly distributed
vs.\ longitude \citep[for example, see ][ for a detailed discussion and a comparison with the capabilities of Doppler Imaging maps]{lan06}. The uniformly distributed component of the spot pattern cannot be derived
because we assume that the unspotted magnitude is that corresponding to the
brighter hemisphere of the star. Only when a long-term sequence of seasonal 
light curves is available can we extract information on the variation
of the uniformly distributed spot pattern from season to season \citep[cf.\ ][]{lan02}.

\subsection{Model parameters}
\label{s:semod_params}

In our spot maps, we consider squared surface map elements of side $s =
18^{\circ}$. while fluxes are always computed with the finer
subdivision $s_{\rm e} = 1^{\circ}$. The stellar parameters
adopted for the modelling of the two objects are listed in Table~\ref{tab:par2}
together with the  reference from which they were taken. 
The spot temperatures are estimated from the $B-V$ and
$V-I$ colour variations according to  \cite{mes06a} and \cite{mes06b}. 

Specifically, we model the amplitude of the {\it V} magnitude versus $B-V$ and $V-I$ 
colour variations induced by the cool/hot spots on the stellar 
photosphere for a grid of spot temperatures and filling factors 
using the Dorren approach \citep{Dorren87} and the
Hauschildt et al.\ \citep{Hauschildt99} atmosphere models. A $\chi^{2}$ minimization is used to select
the solutions that best fit the observed amplitudes.
As expected, the solutions are not unique and
the higher the spot temperature, the higher their
filling factor values. However, the solutions tend to cluster
around a finite range of values for the spot temperature.
Such a finite range of temperature values is adopted as the uncertainty
of our spot temperature determination, which is about $\pm 150$~K
for both stars under analysis.
Assuming that the observed magnitude and colour variations arise from both
cool and hot spots, what we infer from our modelling is an average value
for the temperature inhomogeneities. In the case of HD\,199143 the spot temperature
inferred from either $B-V$ or $V-I$ is in agreement within the computed uncertainty.
For CD$-$64$^{\circ}$1208, the spot
temperature derived from the $B-V$ colour is about 200 K warmer than that derived
from the $V-I$ colour, probably because of the excess flux of 
faculae mainly affecting the B-band \citep{mes06a}. 

\rm
Therefore, the $V-I$ colour probably 
gives a better determination of the cool spot temperature; we estimate an 
uncertainty of $\pm 150$~K for both stars. It is important to note that cool spots
dominate the flux variation of both stars in the V-band whereas the
possible facular component can be regarded as a second order effect in
such extremely active stars \citep[cf.\ the case of HR 1099 as analysed and discussed by][]{vog99}. 
Therefore, the temperature of the cool spots is
the only one appropriate for our modelling of the optical flux
rotational modulation.  From the effective temperatures of the spotted
and the unperturbed photospheres, the spot contrast $C_{\rm s} \equiv
I_{\rm s}/I_{\rm u}$ at the disc centre is evaluated by means of  
Kurucz models, adopting the same mean gravity (see the values listed in
Table~\ref{tab:par2}).
\begin{table}
\begin{center}
\caption{Stellar parameters of HD~199143 and CD$-$64$^{\circ}$1208 adopted for our spot modelling.}
\begin{tabular}{lccc}
\hline 
\hline
 & HD~199143 & CD$-$64$^{\circ}$1208 & Ref. \\
\hline
SpTy & F8V & K8V & 1\\
$M$ ($M_{\odot}$) & 1.50 & 0.80 & 2 \\
$a$ ($R_{\odot}$) & 2.20 & 1.10 & 2 \\
$P_{\rm rot}$ (d) & 0.3558 & 0.3545 & 3 \\
$i$ (deg) & 21.5 & 50.1 & 3 \\
$c/a$ & 0.622 & 0.911 & 3 \\
$T_{\rm eff}$ (K) & 6310 & 4260 & 1\\
$T_{\rm s}$ (K) & 4700 & 3700 & 3 \\
$\log g$ (cm s$^{-2}$) & 3.9 & 4.3 & 2 \\
$u_{V}$ & 0.667 & 0.800 & 3 \\
$C_{\rm s}$ & 0.234 & 0.315 & 3 \\
\hline
\end{tabular}
\label{tab:par2}
\end{center}
~\\
References: 1. Allende Prieto \& Lambert (1999); 2. de La Reza \& Pinz\'on (2004);  3. This work.
\end{table}

The linear limb-darkening coefficients in the V-band, $u_{V}$, are
assumed to be the same for the spotted and  unspotted photospheres,
which does not introduce any significant systematic error given that
the difference between the two coefficients is less than 15\% , and
that the spots occupy a maximum fraction of the order of 10\%\ of the
stellar disc.

The inclination $i$ of the stellar rotation axis with respect to the
line of sight is derived from the observed $v \sin i$, the estimated
stellar equatorial radius $a$ and the rotation period $P_{\rm rot}$
found in Sect.~\ref{sec:timeseries}. Given the errors in the measurements of
$v \sin i$, the error in the inclination is of the order of
$10^{\circ}-15^{\circ}$. Nevertheless, changing the inclination
does not significantly affect the distribution of the relative spotted
area vs.\ longitude that is the primary result of our modelling.

The rapid rotation of the two stars produces a remarkable centrifugal
flattening that makes it necessary to include gravity darkening and
ellipsoidicity in the stellar model.  Any systematic error in the
estimation of such effects modifies the unperturbed flux level of the
star. However, to first order there is no difference in the spotted
area between the most and the least spotted hemispheres, and the
component of the spotted area unevenly distributed in longitude can be
safely mapped.


\section{Results}
\label{sec:results}

The results of our spot modelling are summarized in
Table~\ref{tab:model}. The reduced $\chi^{2}$ is defined as: 
\begin{equation}
\chi^{2} \equiv \frac{1}{M} \sum_{i=1}^{M} \frac{(O_{i} - F_{i})^{2}}{\sigma^{2}},
\label{chisquare}
\end{equation}
where $O_{i}$ is the observed flux and $F_{i}$ the computed flux at
the $i$-th rotational phase. The standard deviation used to compute
the best fit for each light curve has been derived {\it a posteriori} by
considering the standard deviation of the  differences between
the observations and the best fit flux values, both normalized to the maximum observed flux. 
It is comparable with the {\it a priori} standard deviation expected from the accuracy of the
photometry for the first light curve of HD~199143 and the light curve
of CD$-$64$^{\circ}$1208 (i.e.\ 0.005 mag), whereas it is about 2 times smaller
for the second light curve of HD~199143.
\begin{table*}
\caption{Results of the spot modelling of the light curves of HD~199143 and CD$-$64$^{\circ}$1208.
}
\begin{center}
\begin{tabular}{lcccclccccc}
\hline 
\hline
 Star & initial HJD & final HJD & $M$ & $\sigma$ & $\lambda_{\rm ME}$ &  $\lambda_{\rm TR}$ & 
$\chi_{\rm ME}^{2}$ & $\chi_{\rm TR}^{2}$ & $A_{\rm ME}$ & $A_{\rm TR}$ \\
\hline
HD~199143 & 2453555.02 & 2453558.32 & 27 & 0.00620 & 1.0 & 40.0 & 1.1222 & 1.0967 & 0.0200 & 0.0678 \\
HD~199143 & 2453578.34 & 2453580.32 & 63 & 0.00246 & 0.15 & ~6.0 & 1.0366 & 1.0506 & 0.0910 & 0.2392 \\
CD$-$64$^{\circ}$1208 & 2453473.14 & 2453558.26 & 43 & 0.00683 & 2.5 & 50.0 & 0.8542 & 0.8529 & 0.0720 & 0.1620 \\
\hline
\end{tabular}
\end{center}
~\\
\scriptsize{From  left to the right, the name of the star, the initial and final HJD of the light curve, the
number of photometric points in the light curve $M$, their standard deviation $\sigma$, the Lagrangian multiplier of the ME and TR optimizations, the corresponding reduced $\chi^{2}$ and the total
spotted area in units of the entire photosphere of the star, respectively.}
\label{tab:model}
\end{table*}

\begin{figure}[t!]
   \centering
   \includegraphics[angle=90,width=10cm]{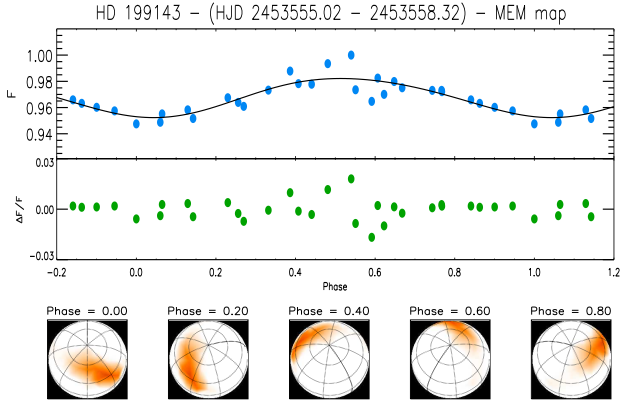}
      \caption{Upper panel: The first light curve of HD~199143 (filled dots) with the corresponding synthetic light curve from the ME best fit spot distribution
(solid line). The flux is given in relative units, assuming as 
the reference value (1.0) the brightest
observation of the star (see Sect.~2). Middle panel: the 
residuals between the observations and the 
model in relative units. Lower panel: The ME spot map of 
the star in orthographic projection, showing the 
disc of the star as it would be 
seen by an observer on the Earth at the labelled rotation phases (see the text).}
         \label{fig3}
   \end{figure}

\begin{figure}[h]
   \centering
   \includegraphics[angle=90,width=10cm]{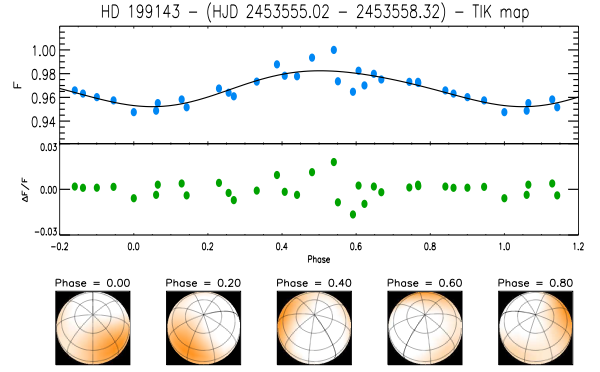}
      \caption{As Fig.~\ref{fig3}, but for the T-regularized best fit.}
         \label{fig4}
   \end{figure}

The best fit of the first light curve of HD~199143 is shown in
Fig.~\ref{fig3}, together with the ME spot map in orthographic
projection as seen by an observer looking at the star from the Earth.
The north pole is always in view because of the small inclination of
the rotation axis of the star. The reference frame is chosen so that the longitude of 
the central meridian of the stellar disc at rotation phase $\phi$
(with $0< \phi <1$) is $360^{\circ} \times \phi$.  The T-regularized
best fit and the corresponding spot maps are shown in Fig.~\ref{fig4}.  The ME- and the T-regularized
best fits of the second light curve of HD~199143 and of the light
curve of CD$-$64$^{\circ}$1208 are shown in Figs~\ref{fig5}-\ref{fig8},
respectively, together with the corresponding spot maps.

\onlfig{7}{
   \begin{figure*}
   \centering
   \includegraphics[angle=90,width=15cm]{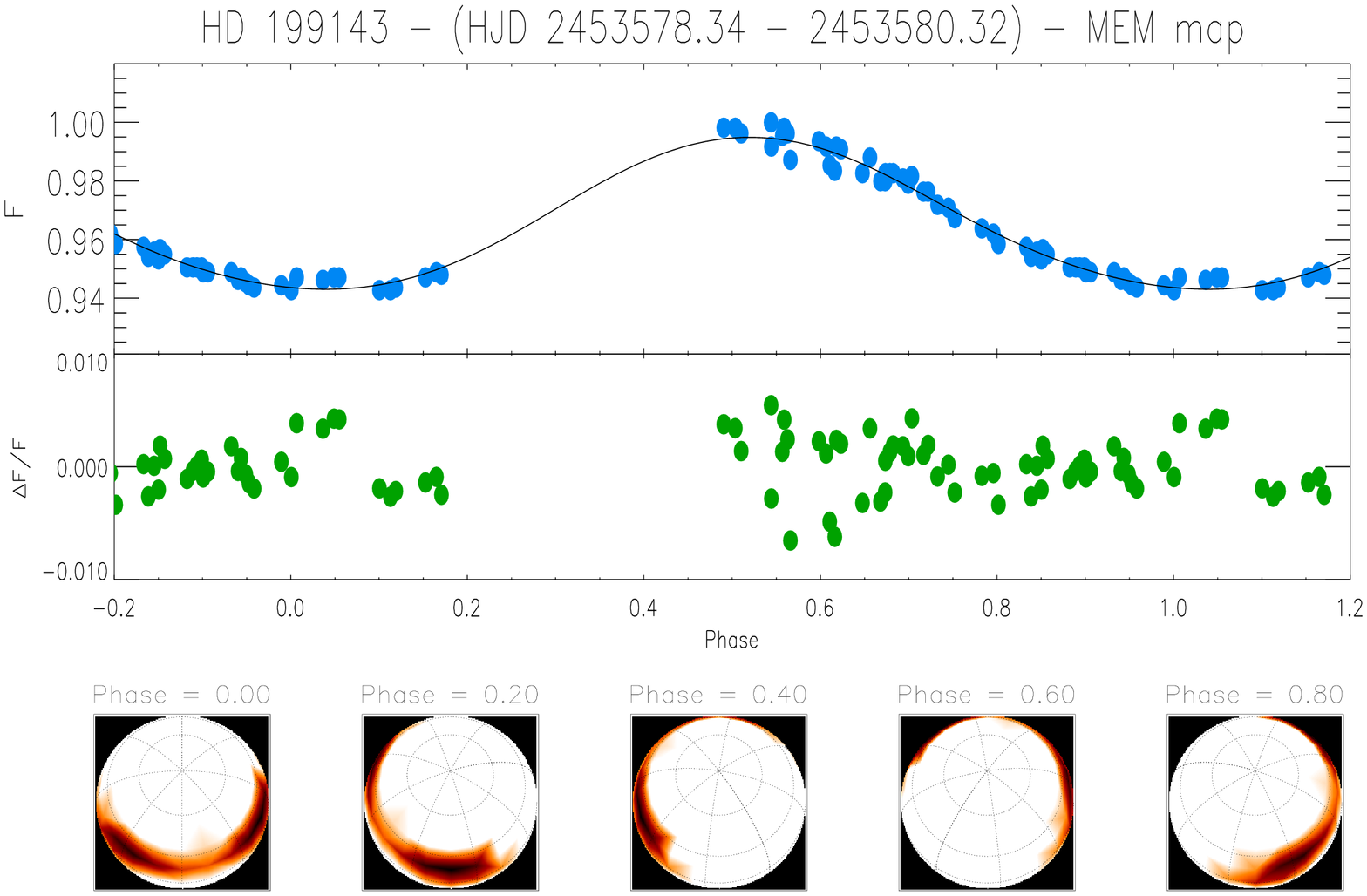}
      \caption{As Fig.~\ref{fig3}, but for the second light curve of HD~199143.}
         \label{fig5}
   \end{figure*}
}
\onlfig{8}{
   \begin{figure*}
   \centering
   \includegraphics[angle=90,width=15cm]{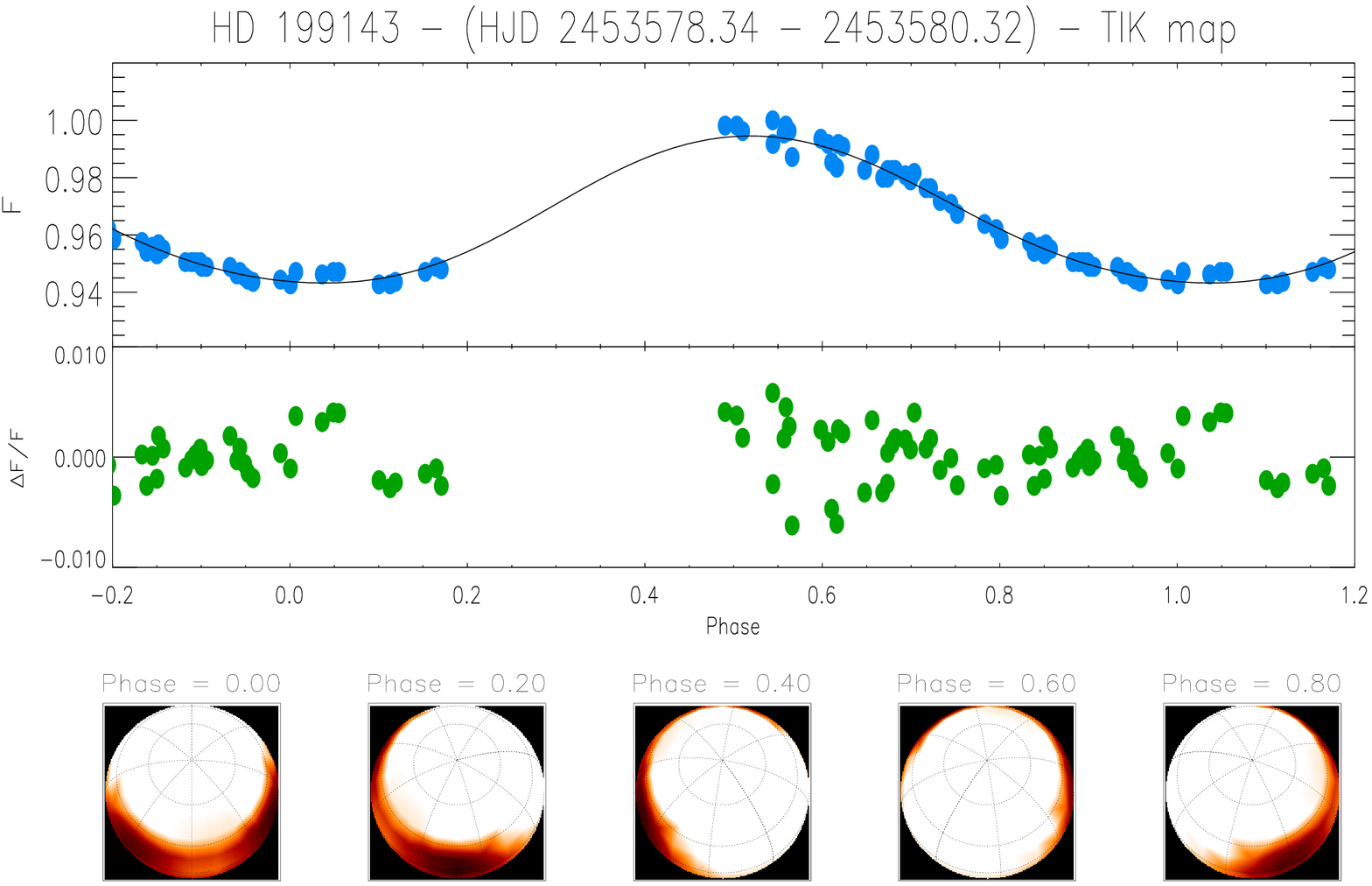}
      \caption{As Fig.~\ref{fig4}, but for the second light curve of HD~199143.}
         \label{fig6}
   \end{figure*}
}
\onlfig{9}{
   \begin{figure*}
   \centering
   \includegraphics[angle=90,width=15cm]{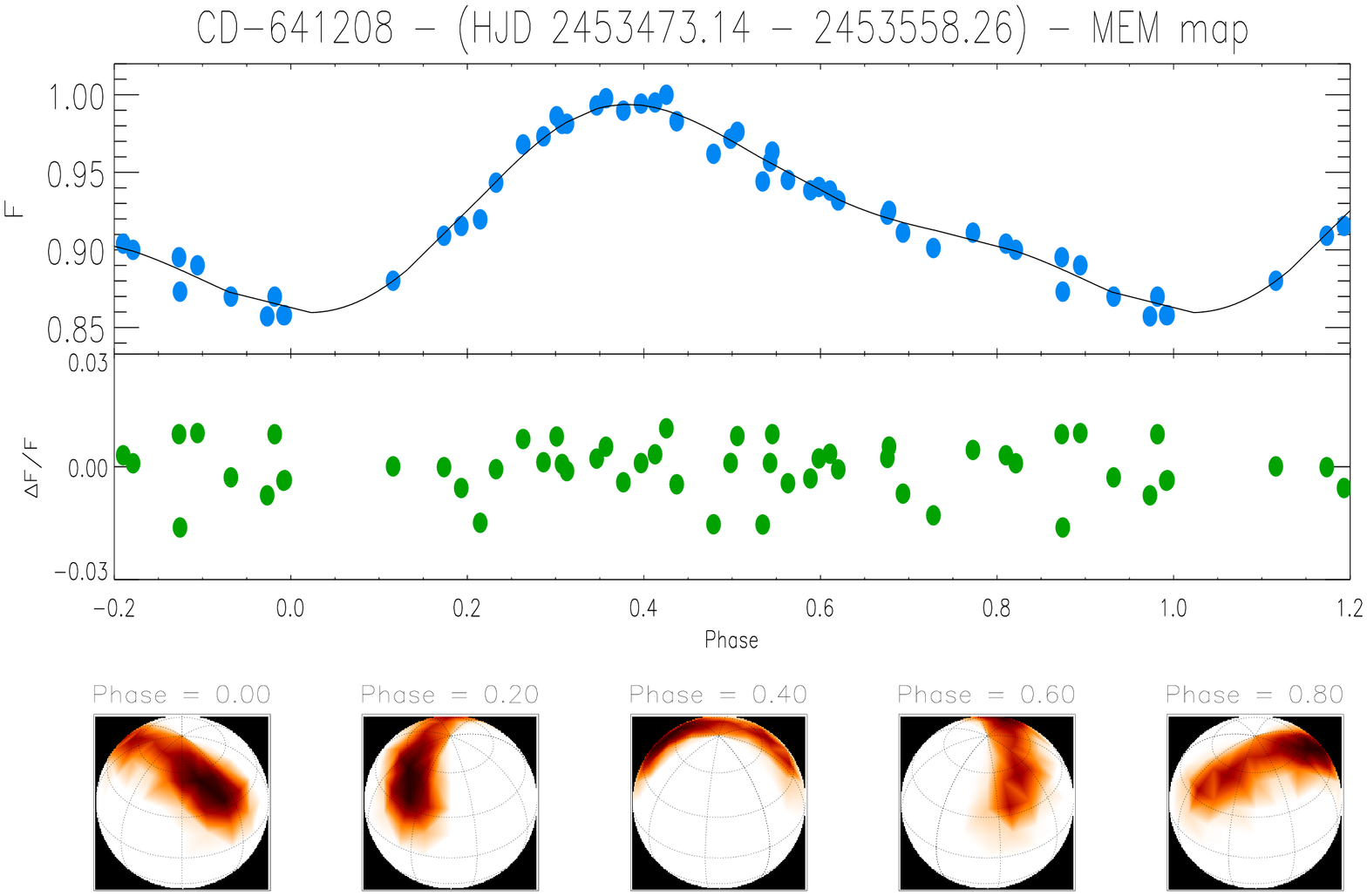}
      \caption{As Fig.~\ref{fig3}, but for the light curve of CD$-$64$^{\circ}$1208.}
         \label{fig7}
   \end{figure*}
}
\onlfig{10}{
   \begin{figure*}
   \centering
   \includegraphics[angle=90,width=15cm]{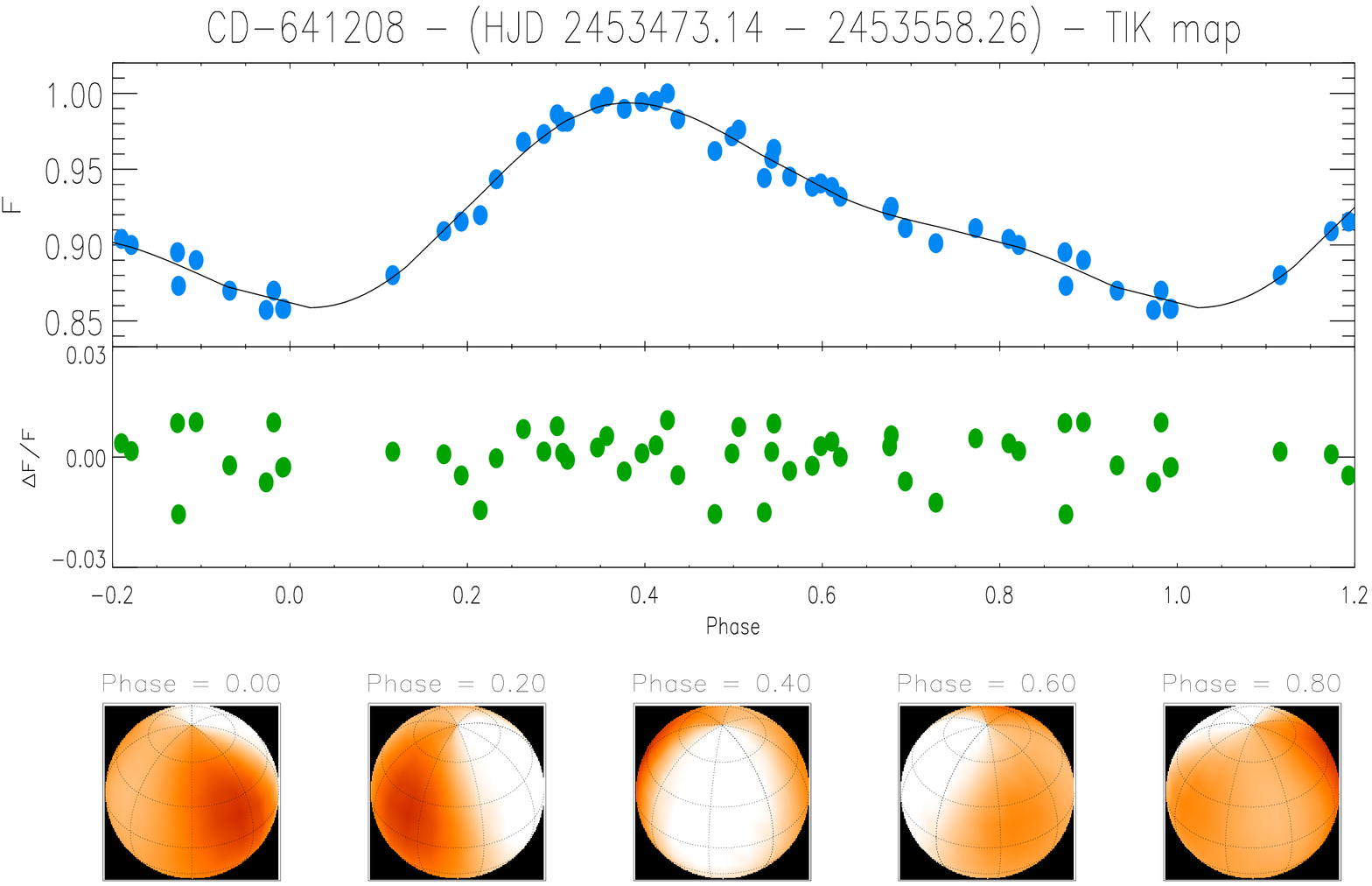}
      \caption{As Fig.~\ref{fig4}, but for the light curve of CD$-$64$^{\circ}$1208.}
         \label{fig8}
   \end{figure*}
}
   \begin{figure}[here]
   \centering
   \includegraphics[width=8cm]{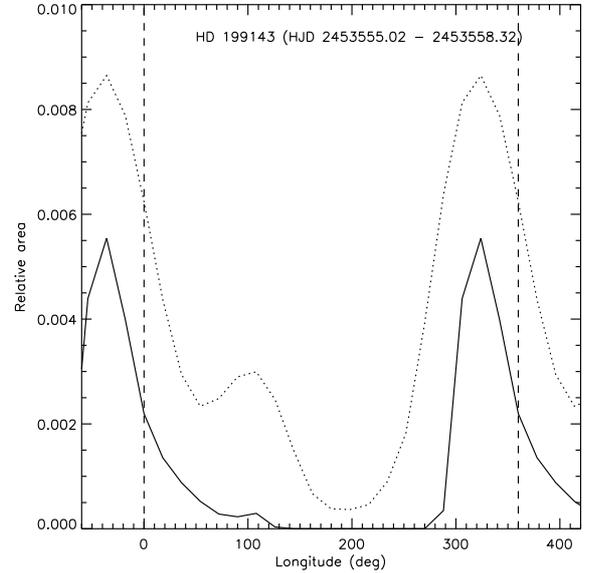}
      \caption{The distribution of the spotted area vs.\ longitude for the ME (solid line) 
and the TR (dotted line) models of the
first light curve of HD~199143. The area unit is the surface of the star. The
distributions have been plotted between $-60^{\circ}$ and $420^{\circ}$ 
to better display 
the structure of the spot pattern; the longitude interval 
$0^{\circ}-360^{\circ}$ is marked by
two vertical dashed lines. }
         \label{fig9}
   \end{figure}

   \begin{figure}[here]
   \centering
   \includegraphics[width=8cm]{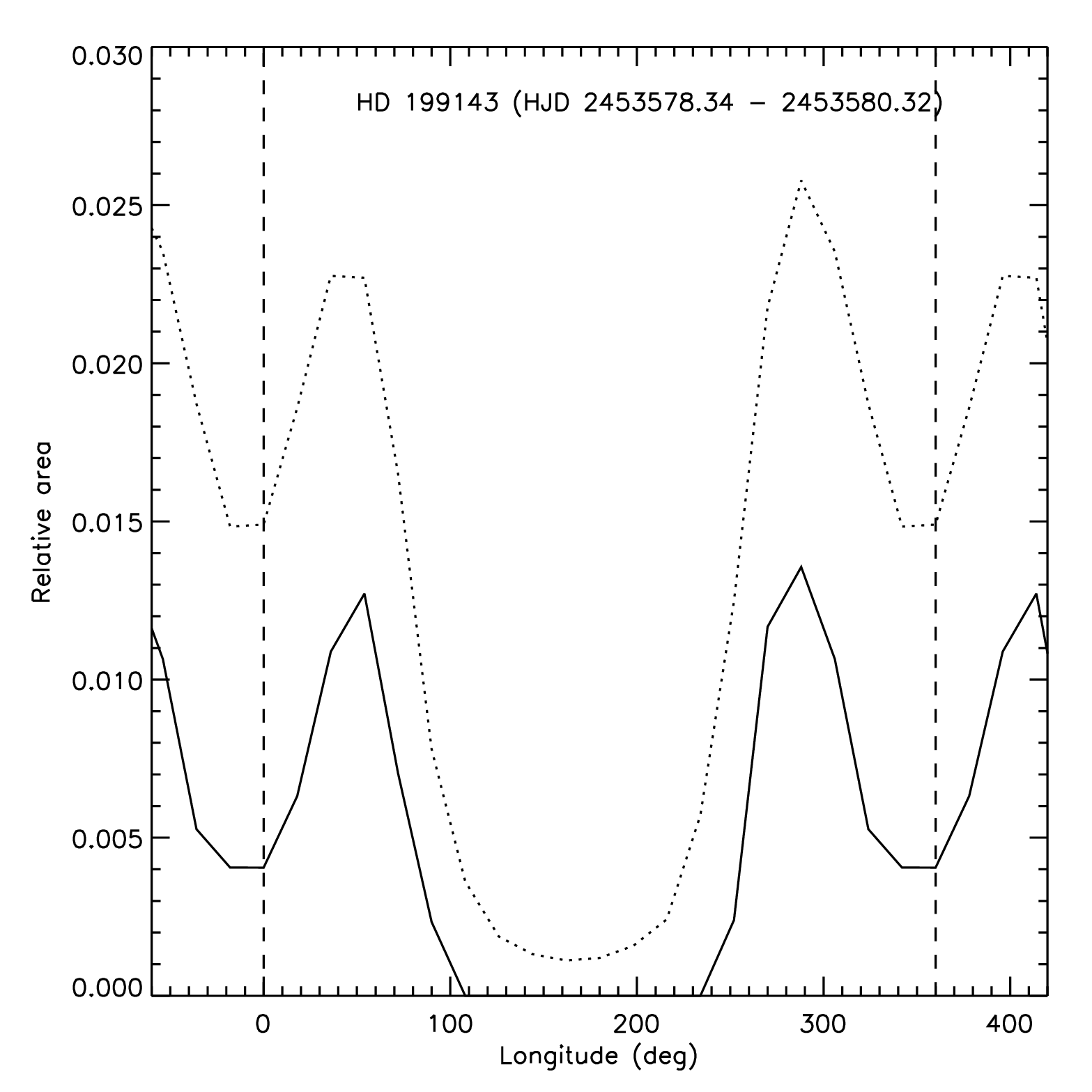}
      \caption{As Fig.~\ref{fig9}, but for the second light curve of CD$-$64$^{\circ}$1208.}
         \label{fig10}
   \end{figure}
   \begin{figure}[here]
   \centering
   \includegraphics[width=8cm]{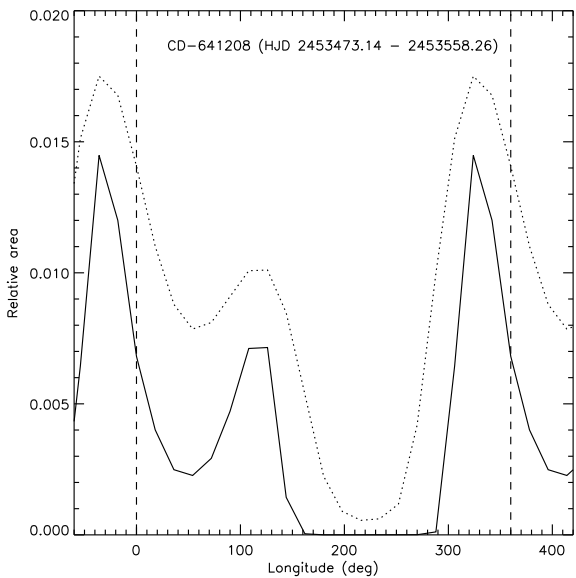}
      \caption{As Fig.~\ref{fig9}, but for the light curve of CD$-$64$^{\circ}$1208.}
         \label{fig11}
   \end{figure}

The spot maps of HD~199143 obtained with the ME and the T
regularizations are remarkably similar, although the TR criterion gives
a smoother distribution of the filling factor, as  expected on
the basis of its {\it a priori} assumptions that select the smoothest
spot distribution that fits the data.  The latitude of the
spots is confined below $\sim 60^{\circ}$ in the first map and below
$\sim 30^{\circ}$ in the second map. This is due to the low
inclination angle of the rotation axis that makes it necessary to put
spots away from the north pole (which is always in view) to produce a
sizeable magnitude variation vs.\ stellar rotation phase. The second light 
curve has a larger amplitude than the first one, 
making it necessary to increase the filling factor {\it and}  put
the spots at lower latitudes to reproduce the observed amplitude. The
lack of spotted area in the ME map between longitudes $100^{\circ}$ and $200^{\circ}$ is due to the lack of  data in the second light curve to constrain the spot pattern in the corresponding phase range. We note the remarkable increase in the total area of the
spot component unevenly distributed in longitude required to fit the
light curve amplitude variation. The ME solutions give an increment of
$\sim 7$\% of the entire photosphere on a time scale of only
$\sim 20$ days.

The spot maps of CD$-$64$^{\circ}$1208 obtained with the ME and the T regularizing
criteria show good longitudinal agreement whereas the latitude
range of the spots is extended to cover the whole visible hemisphere
in the TR map. This is due to the lack of latitudinal constraints on
the spot location due to the higher inclination  of the
stellar rotation axis.  Therefore, in the case of CD$-$64$^{\circ}$1208 what we
can retrieve from the light curve modelling is basically the
distribution of the non-uniform component of the spotted area
vs.\ longitude.

The distributions of the spotted area vs.\ longitude are shown in
Figs~\ref{fig9}-\ref{fig10} for the light curves of HD~199143 and in
Fig.~\ref{fig11} for the light curve of CD$-$64$^{\circ}$1208, respectively.  The
distributions obtained with the ME and TR methods are always very
similar although there is a systematic difference between the ME and
the TR areas, the latter being systematically greater than the
former. This is due to the different assumptions of the two
regularization methods, with ME trying to decrease the spotted area as
much as possible while TR tries to make the spot distribution as
uniform as possible. Therefore, the TR maps have spots in the invisible
part of the stellar surface that systematically increase the total
spotted area in each longitude bin. Of course, the spotted area
projected onto the visible disc, which is responsible for the flux modulation, is the same for both solutions.

The distributions obtained from the first light curve of HD~199143
show the presence of an extended and asymmetric active longitude with
the maximum filling factor at longitude $\sim 325^{\circ}$. A
secondary active longitude is present at $\sim 100^{\circ}$.  It
appears to have grown significantly in the second light curve maps
while the primary active longitude has migrated slightly towards
decreasing longitudes. Unfortunately, the incompleteness of the second
light curve makes such results on the spot pattern evolution quite
uncertain.

{The spot distribution derived for HD~199143 and CD$-$64$^{\circ}$1208 can be compared with those obtained for the two fast-rotating main-sequence stars AB~Dor and LO~Peg. AB~Dor is a K0 dwarf with a rotation period of 
0.5148~days that has been extensively studied through spot modelling \citep[e.g.\  ][]{jarvinenetal05} and Doppler Imaging \citep[e.g.\ ][ and references therein]{jeffersetal07}. Spot modelling based on long-term seasonal photometry generally show one or two active longitudes that change from season to season. In Doppler images, it shows an asymmetric  spotted polar cap of varying extension in addition to several low and intermediate latitude spots.

LO~Peg is a K5-K7 dwarf with a rotation period of only 0.4236~days that has been Doppler imaged by, e.g.,  \cite{barnesetal05}  and \cite{pilusoetal08} showing an asymmetric polar cap with several appendages reaching $10^{\circ}-20^{\circ}$ of latitude. Therefore, the spot distributions obtained from our limited photometric datasets are in general agreement with those obtained for other rapidly rotating young late-type stars. The presence of a more or less asymmetric polar cap can be established only through Doppler imaging because it does not  contribute significantly to the optical flux modulation that is dominated by intermediate and low-latitude spots. Of course, Doppler images have a greater spatial resolution than the maps obtained from the inversion of optical photometry and show that spots as small as a few degrees of longitudes are present on AB~Dor and LO~Peg. Therefore, our active longitudes are likely to be clusters of  much smaller individual spots. }

\section{Conclusions}
\label{sec:conclusions}

We have analysed photometric and high resolution spectroscopic
observations, made during the period 2004--2005, to investigate the
surface activity of the two fastest rotating late-type stars known in
the $\beta$ Pictoris young moving group.  In addition
to surface spot map distributions, these observations have yielded
information on the key physical parameters of rotational velocity,
rotation period, inclinations angle and radial velocity.

\begin{enumerate}

\item Radial velocity standard stars have been used to calculate accurate 
heliocentric radial velocities for each observation by using the
cross-correlation technique. The mean velocities for
HD~199143 and CD$-$64$^{\circ}$1208 are 2.2 $\pm$ 2.9 km~s$^{\rm -1}$ and 2.5
$\pm$ 2.2 km~$s^{\rm -1}$, respectively. These values were used to
calculate the Galactic space-velocity components ($U$,$V$,$W$) and
confirmed their membership of the BPMG.

\item From cross-correlation analysis of our spectra, 
projected rotational velocities, $v \sin i$, of 115.5$\pm$7.5~km~s$^{-1}$ and 121.3$\pm$15.3~km~s$^{-1}$ have been 
measured for HD~199143 and CD$-$64$^{\circ}$1208, respectively.

\item Rotation periods of
0.356$\pm$0.004 days for HD~199143 and 0.355$\pm$0.040 days for CD$-$64$^{\circ}$1208 have been
derived from the time series of V-band differential magnitudes.

\item The inclination, $i$, of the stellar rotation axis with respect 
to the line of sight has been derived from the observed $v \sin i$,
the estimated stellar equatorial radius $a$, and the rotation period
$P_{\rm rot}$. We find inclination angles of $21\fdg 5$ and $50^{\circ}$ with an uncertainty of 
$10^{\circ}-15^{\circ}$ for
HD~199143 and CD$-$64$^{\circ}$1208, respectively.

\item The spot maps of HD~199143 obtained using the TR and ME criteria 
are remarkably similar; the TR criterion gives a smoother distribution
of the filling factor, as expected. The latitude of the spots is
confined below $\sim 60^{\circ}$ in the first map and below $\sim
30^{\circ}$ in the second map. Light curve amplitude variations over
the different epochs of observation require an increase of $\sim 7$\%
in the spot filling factor, unevenly distributed in longitude, on a
time scale of only $\sim 20$ days.

\item The spot maps of CD$-$64$^{\circ}$1208 show good longitudinal agreement
whereas the latitude range of the spots is extended to cover the whole
visible hemisphere in the TR map.

\item The distributions obtained from the first light curve of HD~199143 show the presence of an extended and asymmetric active longitude
with the maximum filling factor at longitude $\sim 325^{\circ}$. A
secondary active longitude is present at $\sim 100^{\circ}$.  It
appears to have grown remarkably in the second light curve maps,
while the primary active longitude has migrated slightly towards decreasing
longitudes. The spotted area distributions on CD$-$64$^{\circ}$1208 show two
active longitudes separated by about $180^{\circ}$ which is not
unusual on such very active stars.
\end{enumerate}

\begin{acknowledgements}
DGA was supported by {\it{Chandra}} grants GO1-2006X and
 GO1-2012X. JJD and VK were supported by NASA contract
 NAS8-39073 to the {\it{Chandra} X-ray Center}. Research at Armagh Observatory is grant-aided by the N.~Ireland Dept. of 
Culture, Arts and Leisure (DCAL). This paper uses observations made at the South African Astronomical Observatory (SAAO) and at the Siding Spring Observatory (SSO), Australia. 

\end{acknowledgements}

\end{document}